\documentclass[12pt,preprint]{aastex}
\def\jcap{JCAP}
\def\beq{\begin{equation}}
\def\eeq{\end{equation}}
\def\ben{\begin{eqnarray}}
\def\een{\end{eqnarray}}
\def\numass{\sum m_{\nu}}
\def\nus{\texttt{MassiveNuS}}
\def\nulcdm{\nu\Lambda{\rm CDM}}
\def\hbj{\hat{\bf J}}
\def\hbe{\hat{\bf e}}
\def\hbt{\hat{\bf T}}

\def\bt{{\bf T}}
\def\hl{\hat{\lambda}}
\def\ev{\,{\rm eV}}
\def\mflip{M_{\rm flip}}
\def\mlflip{\log M_{\rm flip}}
\def\munit{\,h^{-1}\! M_{\odot}}
\def\dunit{\,h^{-1}{\rm Mpc}}

\def\thb{\cos\theta_{2}}
\def\thc{\cos\theta_{3}}
\def\thi{\cos\theta_{i}}
\def\th{\cos\theta}
\def\mctha{\langle\cos\theta_{1}\rangle}
\def\mcthb{\langle\cos\theta_{2}\rangle}
\def\mcthc{\langle\cos\theta_{3}\rangle}

\shortauthors{Lee \& Ryu}
\begin{document}
\title{The Effect of Massive Neutrinos on the Halo Spin Flip Phenomenon}
\author{Jounghun Lee\altaffilmark{1}, Noam I Libeskind\altaffilmark{2,3}, Suho Ryu\altaffilmark{1}}
\altaffiltext{1}{Department of Physics and Astronomy, Seoul National University, Seoul 08826, Republic of Korea  
\email{jounghun@astro.snu.ac.kr, ryu@snu.ac.kr}}
\altaffiltext{2}{Leibniz-Institut f\"{u}r Astrophysik Potsdam (AIP), An der Sternwarte 16, 14482 Potsdam, Germany 
\email{noam@aip.cn.kr}}
\altaffiltext{3}{University of Lyon, UCB Lyon-1/CNRS/IN2P3, IPN Lyon, France}

\begin{abstract}
The halo spin flip refers to the phenomenon that the spin axes of dark matter halos with masses above a certain threshold tend to be preferentially aligned 
perpendicular to the hosting large-scale filaments, while low-mass halos tend to have their spin axes aligned parallel to such structures. Extensive work has so 
far been conducted to understand this phenomenon under the assumption of cold dark matter and suggested that its origin should be closely related to the 
nonlinear evolution of the halo angular momentum in the anisotropic cosmic web. 
We present, for the first time, a numerical examination of this phenomenon assuming the presence of massive neutrinos, finding a clear and robust 
dependence of the threshold mass for the spin flip on the total neutrino mass. 
Our physical explanation is that the presence of more massive neutrinos retard the nonlinear evolution of the cosmic web, which in turn allows the 
halo spin vectors to better retain their memories of the initial tidal interactions in the nonlinear regime.
Our finding implies that the statistical  alignment of halo spins with the large-scale structures can be in principle used as a probe of the total neutrino mass. 
\end{abstract}
\section{Introduction}\label{sec:intro}

The total mass of neutrino species, $\numass$, whose non-zero value was confirmed by the detection of 
neutrino flavor oscillations \citep[for a comprehensive review, see][]{GM08} is of vital importance not only in particle physics but also in cosmology. 
In the former, the non-zero value of $\numass$ is the most conclusive counter proof against the standard model of 
particle physics \citep{GM08}. 
In the latter, the presence of massive neutrinos has an effect of suppressing the growth of the matter densities on a scale determined 
by $\numass$ due to their ability to free stream out of gravitational potential wells \citep{bon-etal80,PS93,PS95}. 
In fact, a close coaction of particle physics with cosmology is required to constrain $\numass$, since laboratory experiments 
have a capacity of putting only a lower limit on $\numass$ \citep{GM08}. The optimal way to determine the 
upper limit of $\numass$, which is most crucial to the physical understanding of their properties, is to resort to the  
cosmological observables that sensitively depend on $\numass$ \citep[for a review, see][]{LP06}. 

The previous works that attempted to probe $\numass$ by using such cosmological observables as the linear density power spectra, 
abundance of galaxy clusters and etc., focused mainly on the suppressing effect of massive neutrinos on the amplitudes of the matter densities 
\citep[e.g., see][and references therein]{LP06,LP12,LP14}.  However, it can be speculated that the presence of massive neutrinos would affect not only 
the amplitudes of the matter densities but also the eigen directions of the tidal fields defined as the second derivative of the gravitational potentials 
and used to quantify the cosmic web \citep{cosmicweb96}. 
A cosmological observable that is susceptible to the effect of massive neutrinos on the tidal eigen directions, if existent and found, may provide a complementary 
probe of $\numass$.  

Here, we attempt to identify one such probe by investigating how the presence of massive neutrinos affects threshold mass at which the preferred 
directions in the spin orientations of dark matter (DM) halos "flip" from parallel to perpendicular to the elongated axes of the surrounding filaments, 
which is often dubbed the "halo spin flip" phenomenon \citep{cod-etal12}. 
The occurrence of the halo spin flip was first witnessed in numerical works based on N-body simulations which investigated the orientations of 
halo spins with respect to surrounding large-scale structures as defined by the eigenvectors of the local tidal tensors and found that the spin vectors of the 
galactic halos having masses lower (higher) than a certain threshold were oriented parallel (perpendicular) to the directions of minimum compression 
\citep[e.g.,][]{ara-etal07,hah-etal07b,paz-etal08,cod-etal12,tro-etal13,lib-etal13,AY14,dub-etal14,for-etal14,cod-etal15a,cod-etal15b,cod-etal18,gan-etal18,
wan-etal18,lee19,kra-etal20}.  
The galaxies resolved in the cosmological hydrodynamic simulations were shown to exhibit weaker signals of the mass-dependent spin flip 
\citep{dub-etal14,cod-etal18,gan-etal19,kra-etal20} than the DM halos, which were attributed to more complicated merging processes of the galaxies 
and baryonic effects.  

What observations found was a signal of the morphology dependent flip of the galaxy spins. While the minor axes of the early-type galaxies tend to be aligned 
with the directions perpendicular to the host filaments, the spin directions of the late-type galaxies exhibit alignments with the directions parallel to the filaments. 
\citep[see also][]{TL13,tem-etal13,pah-etal16,hir-etal17}. 
Very recently, the first observational evidence for the stellar mass dependent  flip of the galaxy spins was reported by \citet{wel-etal20} who utilized 
the data from the Sydney-Australian Astronomical Observatory Multi-object Integral Field Spectrograph surveys \citep{sami} \citep[see also][]{blu-etal19}.

A multitude of scenarios has been put forth to explain what causes the occurrence of the spin flip and why it occurs at a particular threshold mass 
\citep{AY14,wel-etal14,WK18,cod-etal15b}. Although the origin and underlying mechanism has yet to be fully understood, it is now generally accepted that 
the evolutionary process in the cosmic web is largely responsible for the occurrence of the halo spin flip \citep{AY14,wel-etal14,WK18,cod-etal15b}. 
Meanwhile, a recent numerical analysis hinted that the presence of massive neutrinos affects the degree of the anisotropy of the cosmic web \citep{RL20}. 
Given this hint and recalling that the strength and tendency of the tidally induced spin alignments of DM halos depends sensitively on the anisotropy 
of the surrounding web environments \citep{hah-etal07a,lib-etal13,lee19},  we propose a hypothesis that the threshold mass for the halo spin flip may 
also depend on $\numass$. 

The linear tidal field acts on the spin axes of the proto-galactic halos to be aligned with its second eigen direction \citep{whi84,LP00,LP01}.
If the galactic halos became decoupled from the surroundings after the gravitational collapse, their spin directions would retain well 
the initially induced alignments. In reality, the galactic halos located in the cosmic web usually do not become completely decoupled from 
the surroundings but rather prone to their tidal influences. Undergoing the nonlinear evolution, the filamentary cosmic web itself becomes thicker and 
more intricate, whose tidal influence would have an effect of diminishing the strengths of the initially induced alignments of the halo spin 
directions \citep[e.g.,][]{hah-etal10,AY14}. 

The lower-mass galactic halos which form earlier and have sizes smaller than the thickness of the filaments would be more vulnerable to 
the effect of the nonlinearly evolved filamentary cosmic web than the higher-mass ones \citep{AY14}. 
The threshold mass for the spin flip corresponds to the mass scale below which the initially induced alignments of the halo spin axes are 
overwhelmed by the effect of the nonlinearly evolved cosmic web.  The faster the cosmic web evolves, the halo spin flip would occur 
at a higher mass scale.  In the presence of massive neutrinos, the suppressed small-scale powers would retard the nonlinear evolution of the 
cosmic web, which in consequence would lead the spin flip to occur at a lower mass scale. 

Our goal here is to test this hypothesis against N-body simulations performed for the $\nulcdm$ models (neutrinos + cosmological constant $\Lambda$ + cold DM) 
whose initial conditions are different only in $\numass$.  Instead of identifying filamentary structures from the spatial distributions of DM halos, we will directly 
reconstruct the tidal fields from the spatial distributions of the DM particles.  Measuring the alignments between the spin axes of DM halos and the tidal 
eigenvectors, we will explore if and at what mass scale the halo spin directions flip from the second to the third tidal eigenvectors (corresponding to the perpendicular 
to the parallel directions to the filaments). 

Throughout this Paper, we will use the following notations to denote the relevant quantities: ${\bf J}=(J_{i})$ (spin vector of a DM halo), 
$\hat{\bf J}=(\hat{J}_{i})$ (direction of ${\bf J}$), ${\bf T}=(T_{ij})$ (smoothed tidal shear tensor), 
$\hat{\bf T}=(\hat{T}_{ij})$ (traceless version of ${\bf T}$ rescaled by $\vert{\bf T}\vert$), 
$\{\lambda_{i}\}_{i=1}^{3}$ (eigenvalues of ${\bf T}$ in a decreasing order),
$\{{\bf e}_{i}\}_{i=1}^{3}$ (eigenvectors of ${\bf T}$ corresponding to $\{\lambda_{i}\}_{i=1}^{3}$), 
$\{\hat{\bf e}_{i}\}_{i=1}^{3}$ (eigenvectors of $\hat{\bf T}$), $\{\hat{\lambda}_{i}\}_{i=1}^{3}$ (eigenvalues of $\hat{\bf T}$), 
$R_{f}$ (smoothing scale), $M_{h}$ (halo mass), $M_{\rm flip}$ (threshold mass at which the strength of the $\hbj$-$\hbe_{2}$ 
alignment becomes comparable to that of the $\hbj$-$\hbe_{3}$ alignment), and 
$p(\cos\theta_{i})$ (probability density of the cosine of the angle, $\theta_{i}$, between $\hbj$ and $\hbe_{i}$ for $i\in\{1,2,3\}$).

\section{Data and Analysis}\label{sec:analysis}

Our numerical investigation relies entirely on the publicly available data from the Cosmological Massive Neutrino Simulations ($\nus$), 
which is a suite of DM only $N$-body simulations performed on a cosmological box of comoving $512\,h^{-1}$Mpc aside, containing $1024^{3}$ particles 
with individual mass of $10^{10}\,h^{-1}\,M_{\odot}$ \citep{liu-etal18}. A total of $101$ $\nulcdm$ models having unequal initial conditions were adopted 
by the $\nus$ as the background cosmologies, among which three models, with $\numass=0.0,\ 0.1$ and $0.6\ev$, are selected for our analysis 
since they share the same initial conditions other than $\numass$. The analytic linear response approximation was employed by \citet{liu-etal18} 
to include the massive neutrinos in the background for the $\nus$. 

The $\nus$ also provides a catalog of bound objects identified by the Rockstar algorithm \citep{rockstar}, which includes not only the distinct 
halos but also their substructures. Eliminating the substructures from the catalog, we select the distinct galactic halos with masses in the logarithmic range of 
$11.8\le \log (M_{h}/\munit)<13$, for each of the three selected models.
Table \ref{tab:initial} lists the values of the key cosmological parameters and the numbers of the distinct galactic halos ($N_{g}$) for the three selected models. 
Note that although the three models share the same value of the primordial power spectrum amplitude ($A_{s}$), they differ from one another in the value of the 
rms density fluctuation within a top-hat radius of $8\dunit$ ($\sigma_{8}$).  

Dividing the simulation box into a grid of $256^{3}$ cells, we determine the raw density contrast, $\delta({\bf x})$, at the location of each grid cell, ${\bf x}$, 
by applying a cloud-in-cell algorithm to the particle distribution at $z=0$. 
Performing the Fast Fourier Transformation (FFT) of $\delta({\bf x})$, we obtain its Fourier amplitude, $\tilde{\delta}({\bf k})$, 
at each Fourier-space wave vector, ${\bf k}=(k\hat{k}_{i})$. 
An inverse FFT of $\tilde{T}_{ij}({\bf k})\equiv \hat{k}_{i}\hat{k}_{j}\tilde{\delta}({\bf k})\exp\left[-k^{2}R^{2}_{f}/2\right]$ 
returns, $T_{ij}({\bf x})$ with $i,j\in\{1,2,3\}$,  the tidal field smoothed by a Gaussian window function on the scale of $R_{f}$. 

Locating the grid point, ${\bf x}_{h}$, where each galactic halo resides, we calculate $\hbt({\bf x}_{h})$ by subtracting the trace from $\bt({\bf x}_{h})$ 
and rescaling it by its magnitude. Finding $\{\hl\}_{i=1}^{3}$ and $\{\hbe\}_{i=1}^{3}$ at the location of each galactic halo through a similarity transformation 
of $\hbt({\bf x}_{h})$, we compute the projection of $\hbj$ onto each tidal eigenvector as $\cos\theta_{i}=\vert\hbj\cdot\hbe_{i}\vert$. 
Splitting the logarithmic mass range, $11.8\le \log M_{h}<13$, into six differential bins, we determine the probability density distribution, 
$p(\thi)$, as well as the ensemble average at each mass bin.
If $\hbj$ is not aligned with $\hbe_{i}$,  we expect a uniform distribution of $p(\cos\theta_{i})=1$. 
If $\hbj$ is aligned with the direction parallel (perpendicular) to $\hbe_{i}$, we expect $p(\cos\theta_{i})$ to be an increasing (decreasing) function of 
$\cos\theta_{i}$, yielding $\langle\cos\theta_{i}\rangle>0.5$ ($\langle\cos\theta_{i}\rangle<0.5$). 
The linear tidal torque theory (TTT) \citep{whi84} predicts $\mcthb>0.5$, $\mcthc\sim 0.5$ and $\mctha<0.5$ in the proto-galactic stages, 
regardless of $M_{h}$ \citep{LP00}. 

\begin{figure}
\begin{center}
\includegraphics[scale=0.7]{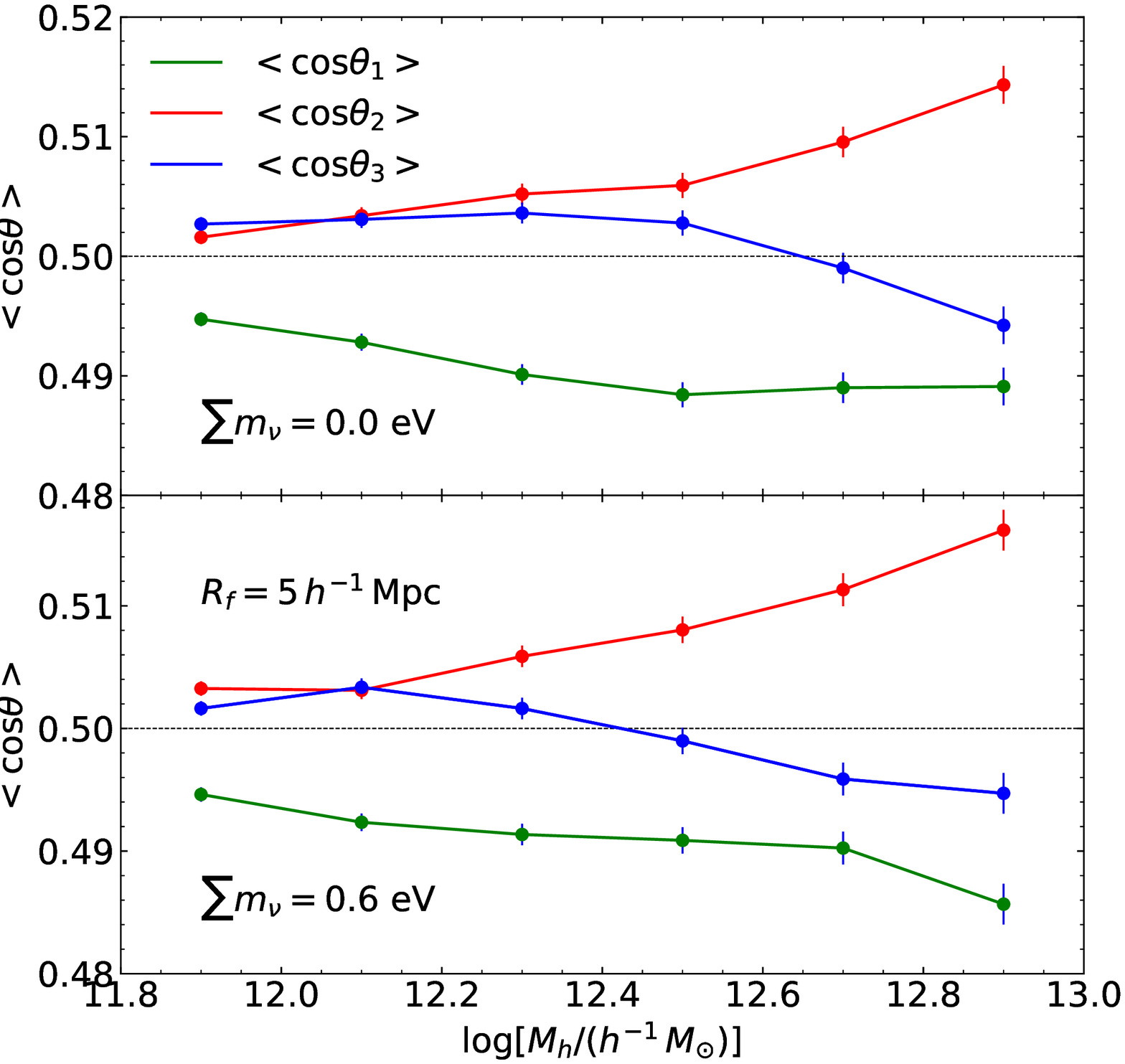}
\caption{Mean values of the cosines of the angles between the halo spin vectors and each of three eigenvectors of the local tidal field 
smoothed on the scale of $R_{f}=5\,h^{-1}$Mpc as a function of the halo mass $M_{h}$ for two different cases of the total neutrino mass 
$\numass$. The dotted line in each panel corresponds to the case of random spin orientations, $\langle\cos\theta_{i}\rangle=0.5$.}
\label{fig:eali_5}
\end{center}
\end{figure}

Figure \ref{fig:eali_5} plots $\mctha$ (green lines), $\mcthb$ (red lines)  and $\mcthc$ (blue lines) at the six logarithmic mass bins for the two cases of 
$\numass=0.0\ev$ (top panel) and $\numass=0.6\ev$ (bottom panel). For this plot, we set $R_{f}$ at $5\,h^{-1}$Mpc,  leaving out the 
results for the case of $\numass=0.1\ev$, which turns out to be almost the same as those for the case of $\numass=0.0\ev$. The errors are calculated 
as one standard deviation in the mean value as $\left[\left(\langle\cos^{2}\theta_{i}\rangle-\langle\cos\theta_{i}\rangle^{2}\right)/\left(n_{g}-1\right)\right]^{1/2}$, 
where $n_{g}$ denotes the number of the distinct DM halos in each bin. 
As can be seen, the two $\nulcdm$ models yield a similar trend.  As $M_{h}$ decreases, the value of $\mcthb$ almost monotonically diminishes down to 
$0.5$, while the values of $\mctha$ and $\mcthc$ mildly increase. In the entire mass range of $11.8\le \log M_{h}<13$, 
the value of $\mcthb$ ($\mctha$) remains higher (lower) than $0.5$. Whereas, the value of ($\mcthc-0.5$) switches its sign midway,  which leads 
$\mcthc\ge\mcthb$ in the mass range below a certain threshold. 
The two $\nulcdm$ differ from each other in the rate at which $\mcthc$ increases with the decrement of $M_{h}$ and in the value of the threshold mass 
at which $\mcthc\sim\mcthb$. 

We define, $\mflip$, as the threshold mass at which the strength of the {\it parallel} alignment between $\hbj$ and $\hbe_{3}$ becomes comparable to that 
between $\hbj$ and $\hbe_{2}$. To find $\mflip$ for the two $\nulcdm$ models, we statistically evaluate the similarity between the strengths of the $\hbj$-$\hbe_{2}$ 
and $\hbj$-$\hbe_{3}$ alignments at each bin.  Instead of comparing simply $\mcthc$ with $\mcthb$, we take a more rigorous approach, performing the 
Kolmogorov–Smirnov (KS) test of the null hypothesis of $p(\thb)\sim p(\thc)$. If the spin flip occur at a certain mass bin, then the confidence level for the 
rejection of this null hypothesis by the KS test would drop below $99.9\%$. 

It is worth mentioning here the advantage of defining $\mflip$ as a threshold mass at which $p(\thc)\sim p(\thb)$ and $\mcthc\ge 0.5$.  The previous works 
conventionally defined $\mflip$ as the threshold mass at which $\mcthc\ge 0.5$. 
However, this conventional definition of $\mflip$ does not take into proper account the possibility that $\hbj$ can be simultaneously aligned with both of 
$\hbe_{2}$ and $\hbe_{3}$ (i.e., $\mcthb>0.5$ and $\mcthc>0.5$).  If the $\hbj$-$\hbe_{2}$ alignment is stronger than the $\hbj$-$\hbe_{3}$ alignment 
(i.e., $\mcthb>\mcthc>0.5$), then $\hbj$ would appear to be aligned perpendicular to the elongated axes of the filaments (i.e., the directions of minimum 
compression) in spite of $\mcthc>0.5$. The neglect of this possibility would result in a spurious value of $\mflip$. 
Suppose that $\mcthb>\mcthc>0.5$ at a given mass $M_{h}$. According to our definition, we would properly conclude 
$\mflip<M_{h}$, while the conventional method based only on $\mcthc$ would spuriously claim $\mflip>M_{h}$. 

Two cumulative distributions, $P(\thb<\th)$ and $P(\thc<\th)$ defined as $P(\thi<\th)\equiv \int_{0}^{\th}\,d\cos\theta^{\prime}_{i}\,p(\cos\theta^{\prime}_{i})$, 
are determined. If there is no alignment between $\hbj$ and $\hbe_{i}$, we expect $P(\thi<\th)=\th$. The alignment of $\hbj$ with the parallel and perpendicular 
directions  of $\hbe_{i}$ would yield $P(\thi<\th)<\th$ and $P(\thi<\th)>\th$, respectively. 
We calculate the maximum distance between the two distributions at each mass bin as 
\begin{equation}
\label{eqn:ks_max}
D_{2,3}=\max\,\vert P(\thc<\th)-P(\thb<\th)\vert\, ,
\end{equation}
and multiply $D_{2,3}$ by $\sqrt{n_{g}/2}$ where $n_{g}$ is the number of the galactic halos at a given mass bin. If this quantity, $\sqrt{n_{g}/2}\,D_{2,3}$, 
is larger than $1.949$, then the null hypothesis is rejected at the confidence level higher than $99.9\%$. 
 
\begin{figure}
\begin{center}
\includegraphics[scale=0.7]{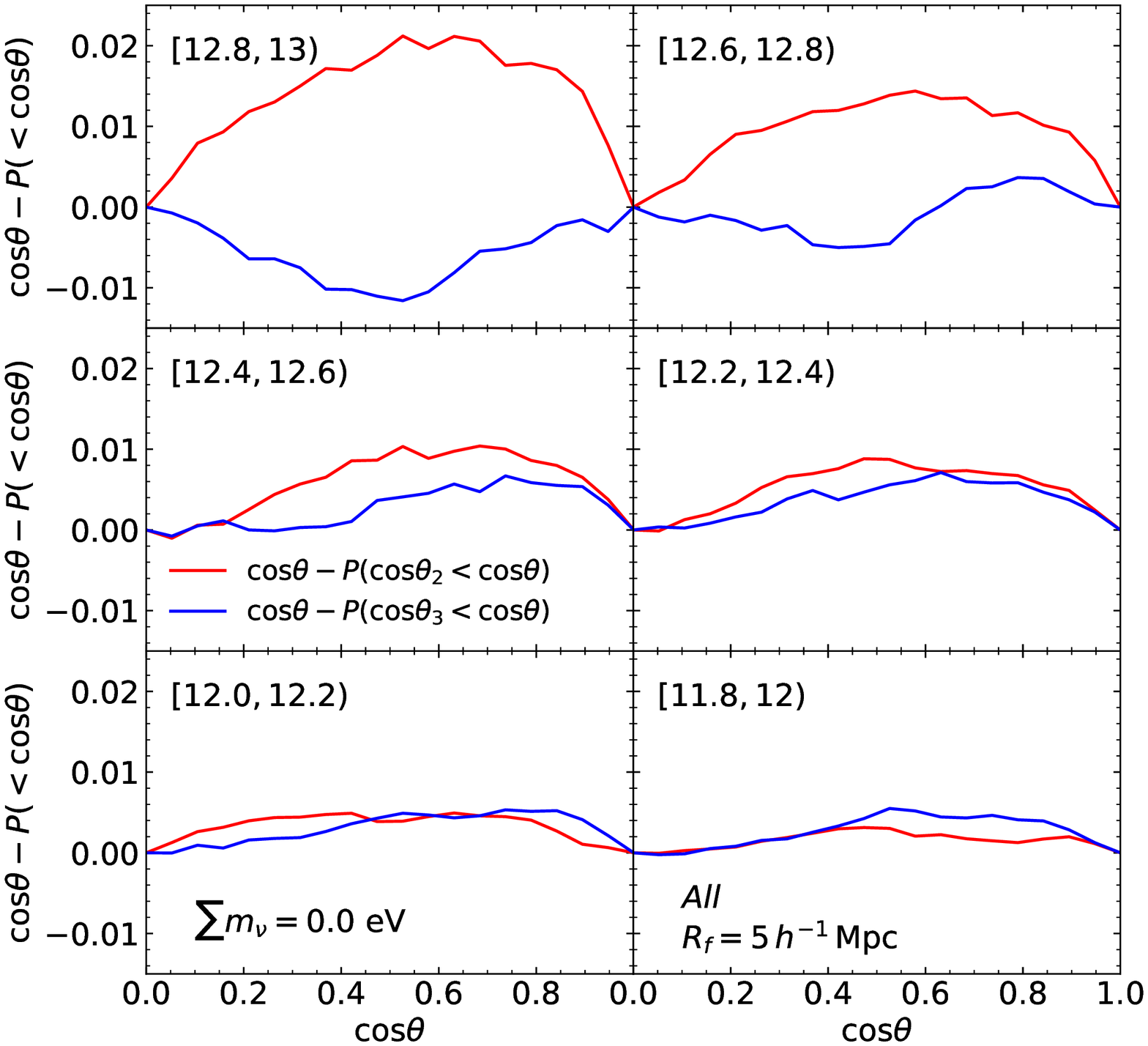}
\caption{Differences between $\th$ and $P(\thi<\th)$ with $i\in\{2,3\}$ for the case of $\numass=0.0\ev$ 
at six different logarithmic mass bins of $12.8\le \log M_{h}< 13,\ 12.6\le \log M_{h}<12.8,\ 
12.4\le \log M_{h}<12.6,\ 12.2\le \log M_{h}< 12.4,\ 12.0\le \log M_{h}<12.2,\ 11.8\le \log M_{h}<12.0$ 
in the top-left, top-right, middle-left, middle-right, bottom-left, and bottom-right panels, respectively.}
\label{fig:cbin_all_5_0.0}
\end{center}
\end{figure}

\begin{figure}
\begin{center}
\includegraphics[scale=0.7]{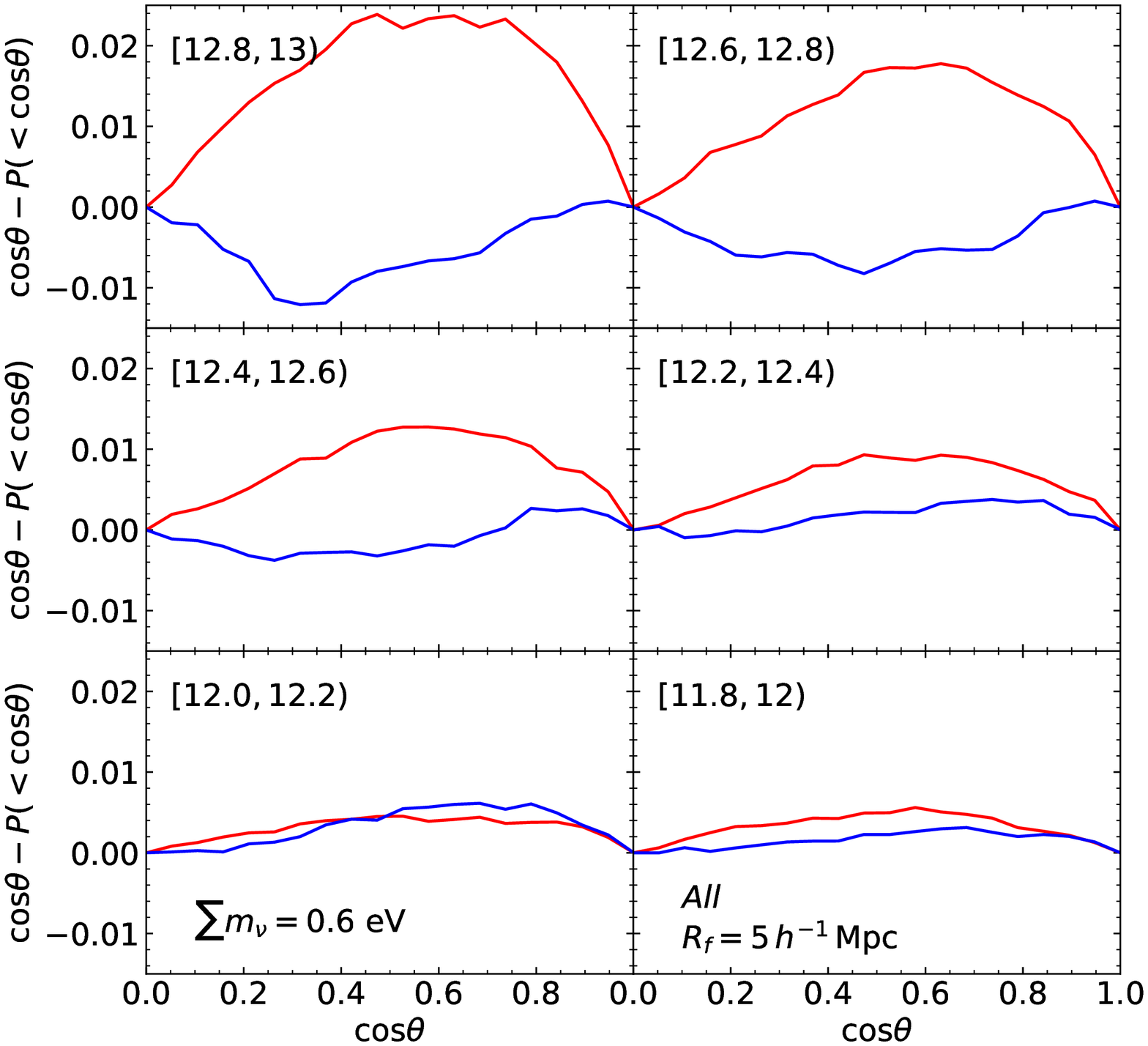}
\caption{Same as Figure \ref{fig:cbin_all_5_0.6} but for the case of $\numass=0.6\ev$.}
\label{fig:cbin_all_5_0.6}
\end{center}
\end{figure}

The six panels of Figure \ref{fig:cbin_all_5_0.0} show $\th-P(\thb<\th)$ (red lines) and $\th-P(\thc<\th)$ (blue lines) for the case of 
$\numass=0.0\ev$ at the six logarithmic mass bins:  
$12.8\le \log M_{h}< 13$ (top left panel), $12.6\le \log M_{h}< 12.8$ (top right panel), $12.4\le \log M_{h}< 12.6$ (middle left panel), 
$12.2\le \log M_{h}< 12.4$ (middle right panel), $12.0\le \log M_{h}< 12.2$ (bottom left panel), and $11.8\le \log M_{h}< 12$ (bottom right panel). 
The KS test rejects the null hypothesis at the confidence levels higher than $99.9\%$ in the first three bins but only at the $90\%$ confidence 
level at the fourth mass bin (middle right panel) where $\mcthc>0.5$, indicating the occurrence of the spin flip in the fourth mass bin, i.e., 
$\mlflip\sim (12.3\pm 0.1)$ for the case of $\numass=0.0\ev$. 
Figure \ref{fig:cbin_all_5_0.6} shows the same as Figure \ref{fig:cbin_all_5_0.0} but for the case of $\numass=0.6\ev$, revealing that the null hypothesis is 
rejected at the $99.9\%$ confidence level in the fourth mass bin unlike the case of $\numass=0.0\ev$. The drop of the confidence level below $99.9\%$ 
occurs at the fifth mass bin (bottom left panel) where $\mcthc>0.5$, indicating $\mlflip \sim (12.1\pm 0.1)$ for the case of $\numass=0.6\ev$. 
These results imply that the presence of more massive neutrinos has an effect of rendering the spin flip to occur at lower mass scales. 

\begin{figure}
\begin{center}
\includegraphics[scale=0.7]{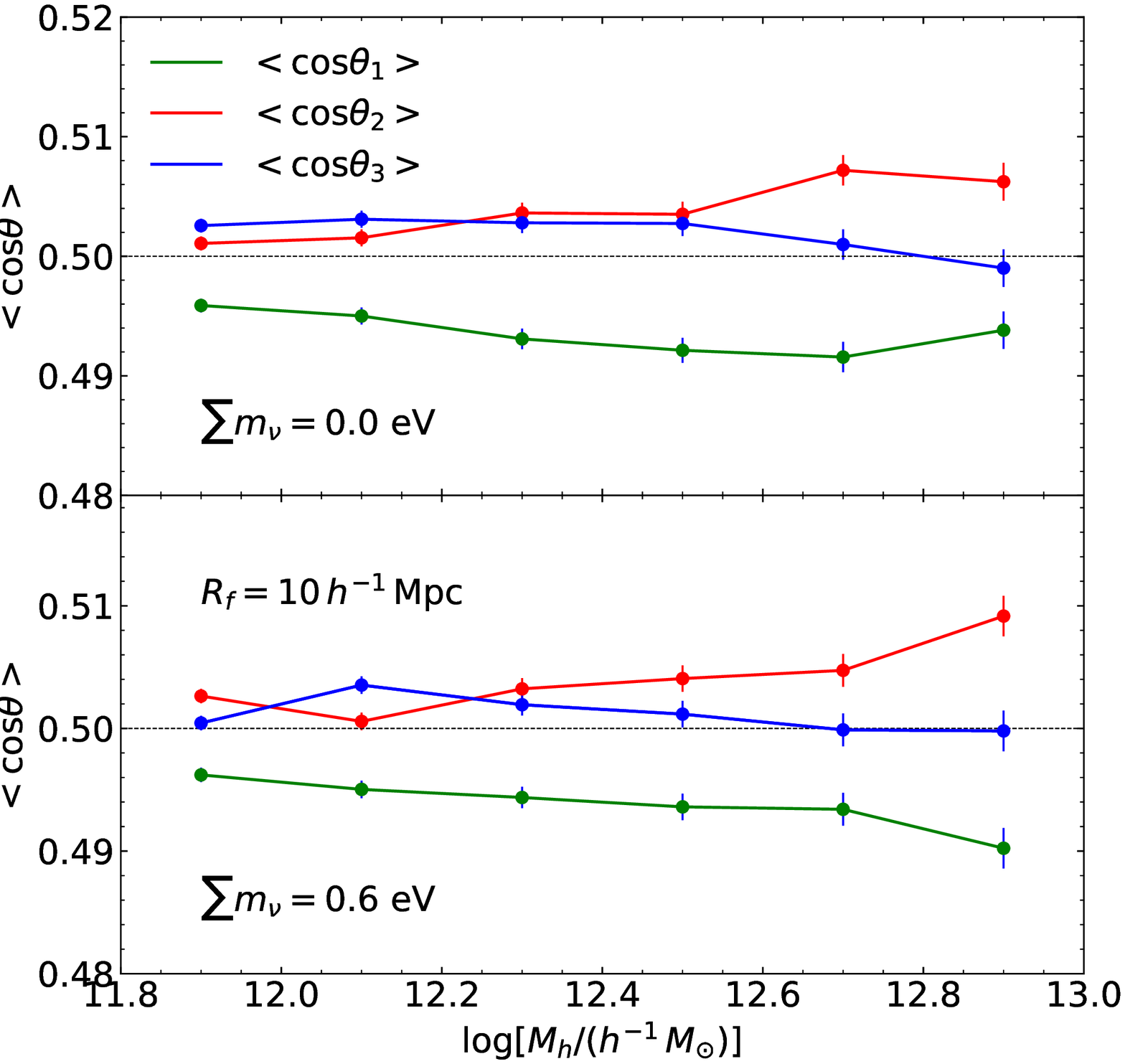}
\caption{Same as Figure \ref{fig:eali_5} but on the scale of $R_{f}=10\dunit$.}
\label{fig:eali_10}
\end{center}
\end{figure}

\begin{figure}
\begin{center}
\includegraphics[scale=0.7]{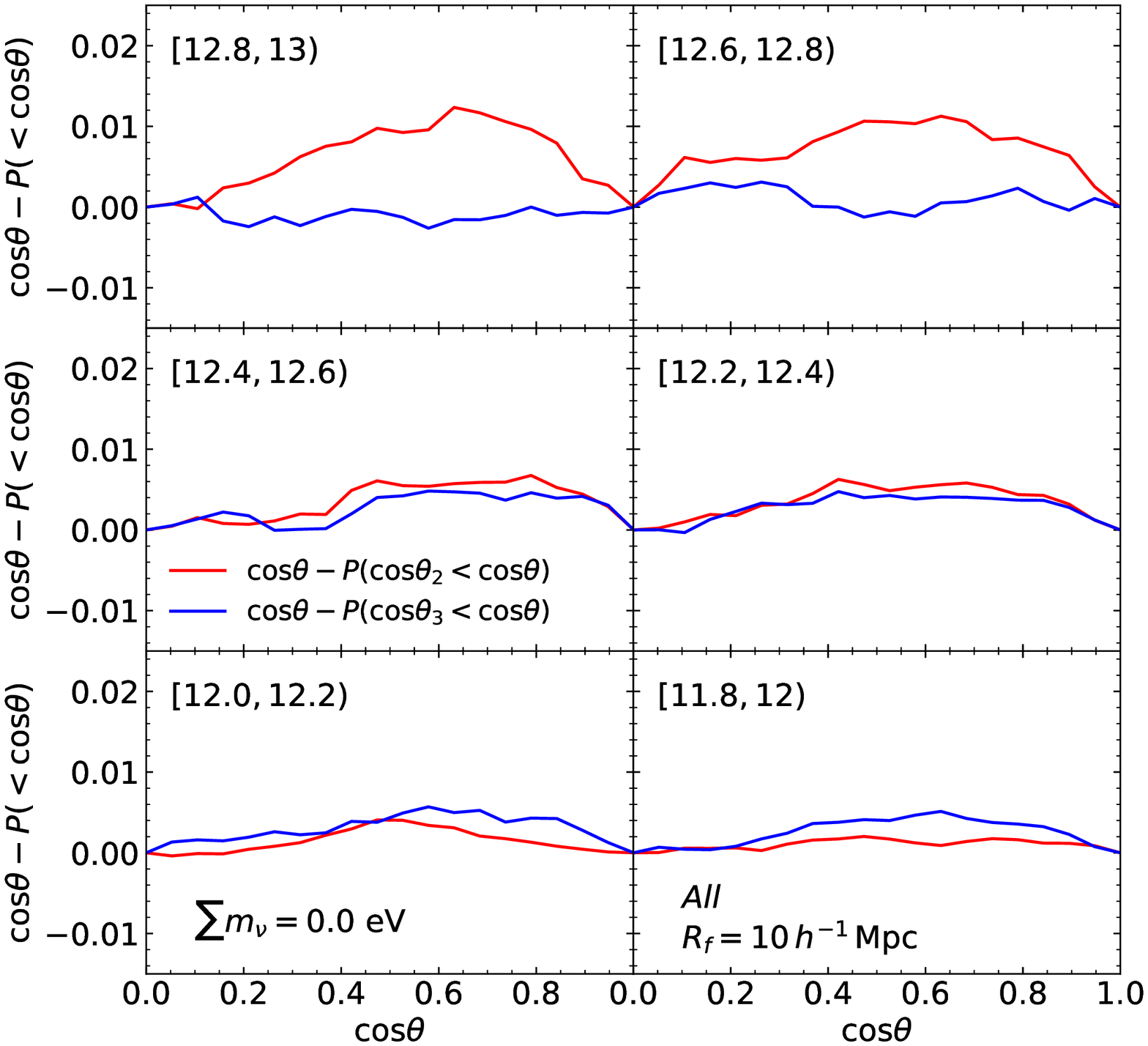}
\caption{Same as Figure \ref{fig:cbin_all_5_0.0} but on the scale of $R_{f}=10\dunit$.}
\label{fig:cbin_all_10_0.0}
\end{center}
\end{figure}

\begin{figure}
\begin{center}
\includegraphics[scale=0.7]{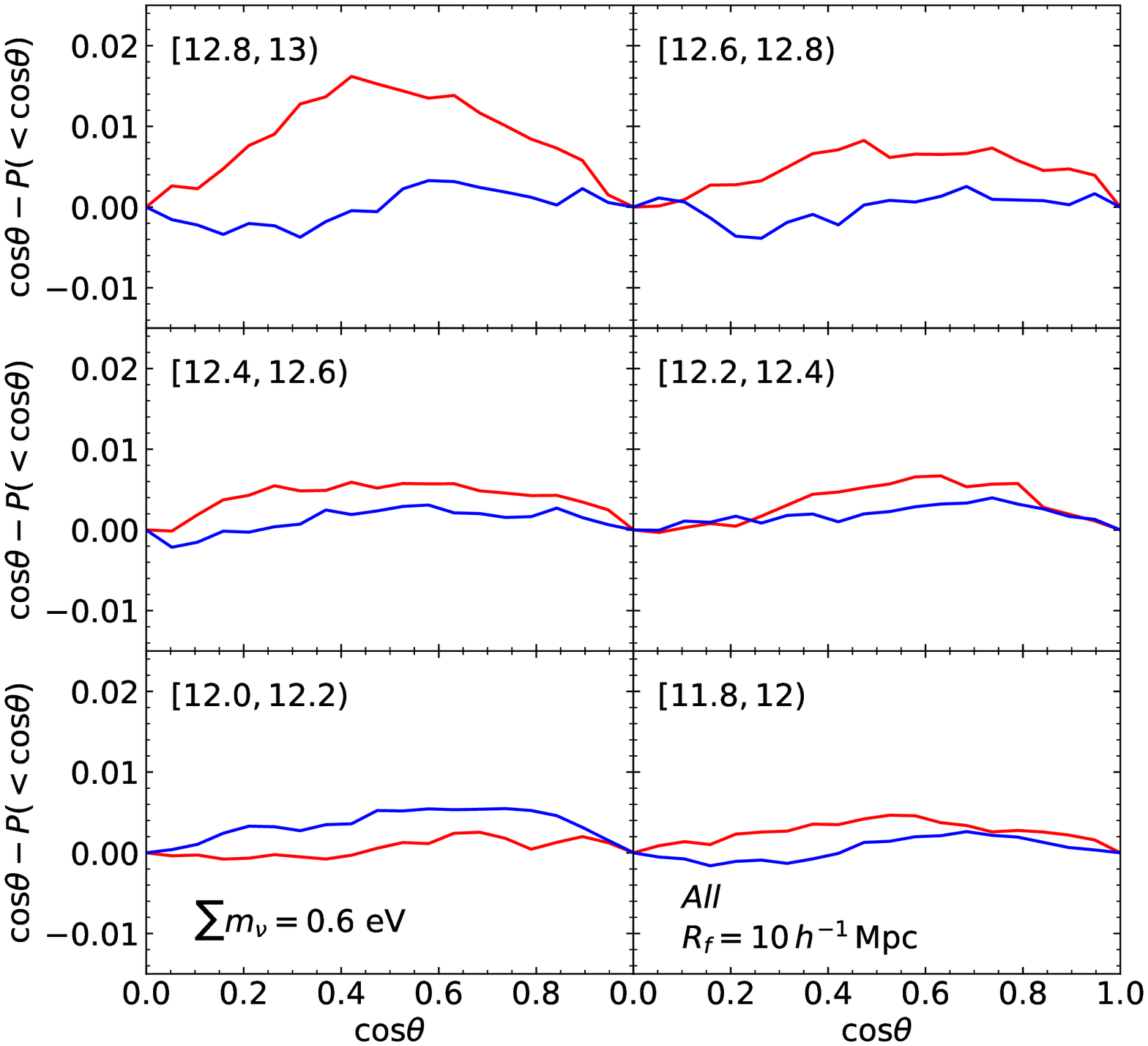}
\caption{Same as Figure \ref{fig:cbin_all_5_0.6} but on the scale of $R_{f}=10\dunit$.}
\label{fig:cbin_all_10_0.6}
\end{center}
\end{figure}

To see whether or not our detection of the dependence of $\mflip$ on $\numass$ is robust against the variation of $R_{f}$, we smooth $\bt$ on the larger scale 
of $R_{f}=10\dunit$ and repeat the whole process, the results of which are displayed in Figures \ref{fig:eali_10}-\ref{fig:cbin_all_10_0.6}. As can be seen, 
the increase of $R_{f}$ weakens the overall $\hbj$-$\hbe_{i}$ alignments, which is consistent with the previous works \citep[e.g.,][]{TL13,lee19} . It also 
leads the spin flip phenomenon to occur at a larger mass bin for both of the $\nulcdm$ models.  The drop of the confidence level for 
the rejection of the null hypothesis below $99.9\%$  is found in the third mass bin of $\mlflip\sim (12.5\pm 0.1)$ for the case of $\numass=0.0\ev$ and 
in the fourth mass bin of $\mlflip\sim (12.3\pm 0.1)$ for the case of $\numass=0.6\ev$. This result confirms that $\mflip$ depends on 
$\numass$, regardless of $R_{f}$.  

\begin{figure}
\begin{center}
\includegraphics[scale=0.7]{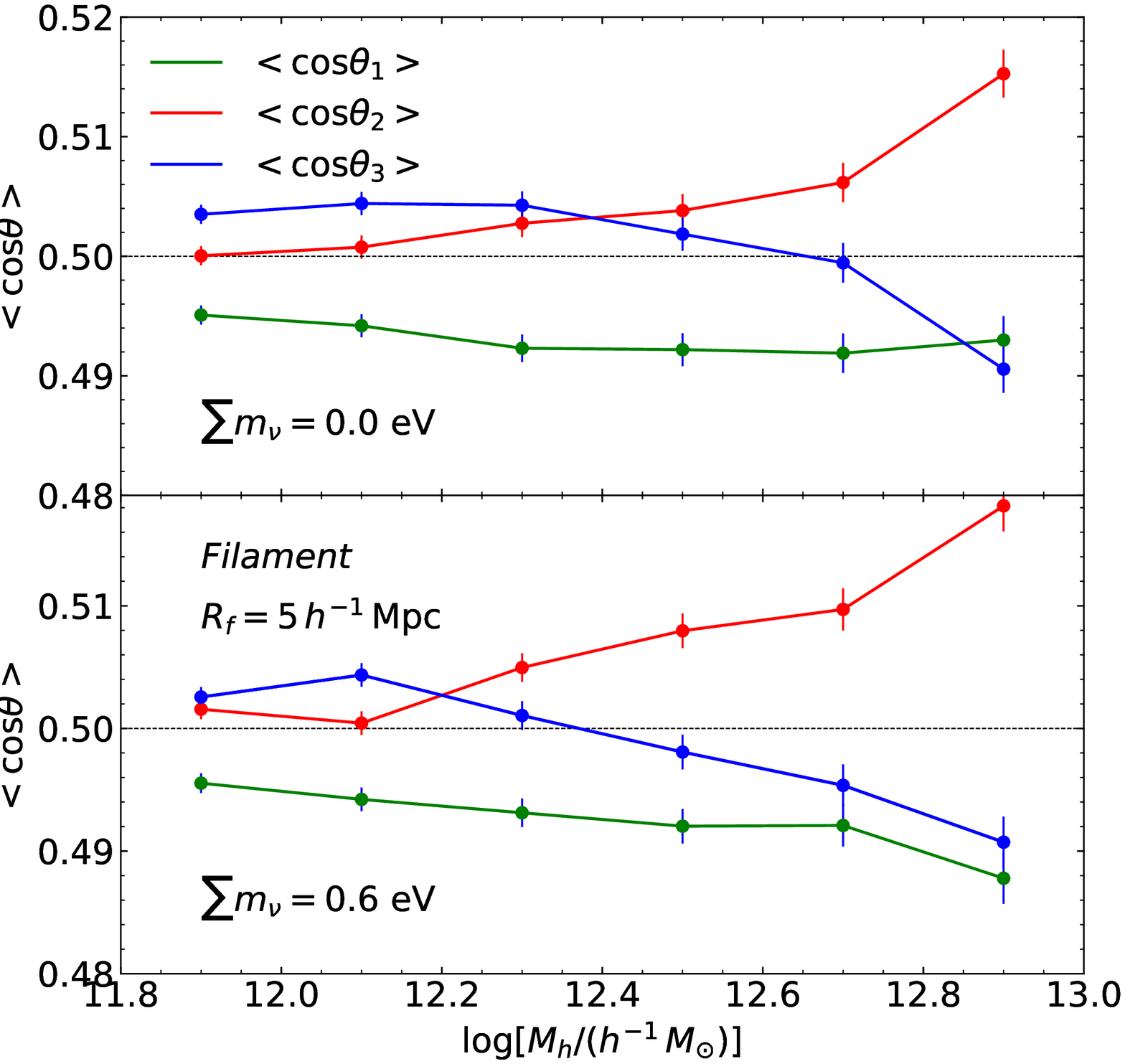}
\caption{Same as Figure \ref{fig:eali_5} but in the filament environment.}
\label{fig:eali_fil_5}
\end{center}
\end{figure}

\begin{figure}
\begin{center}
\includegraphics[scale=0.7]{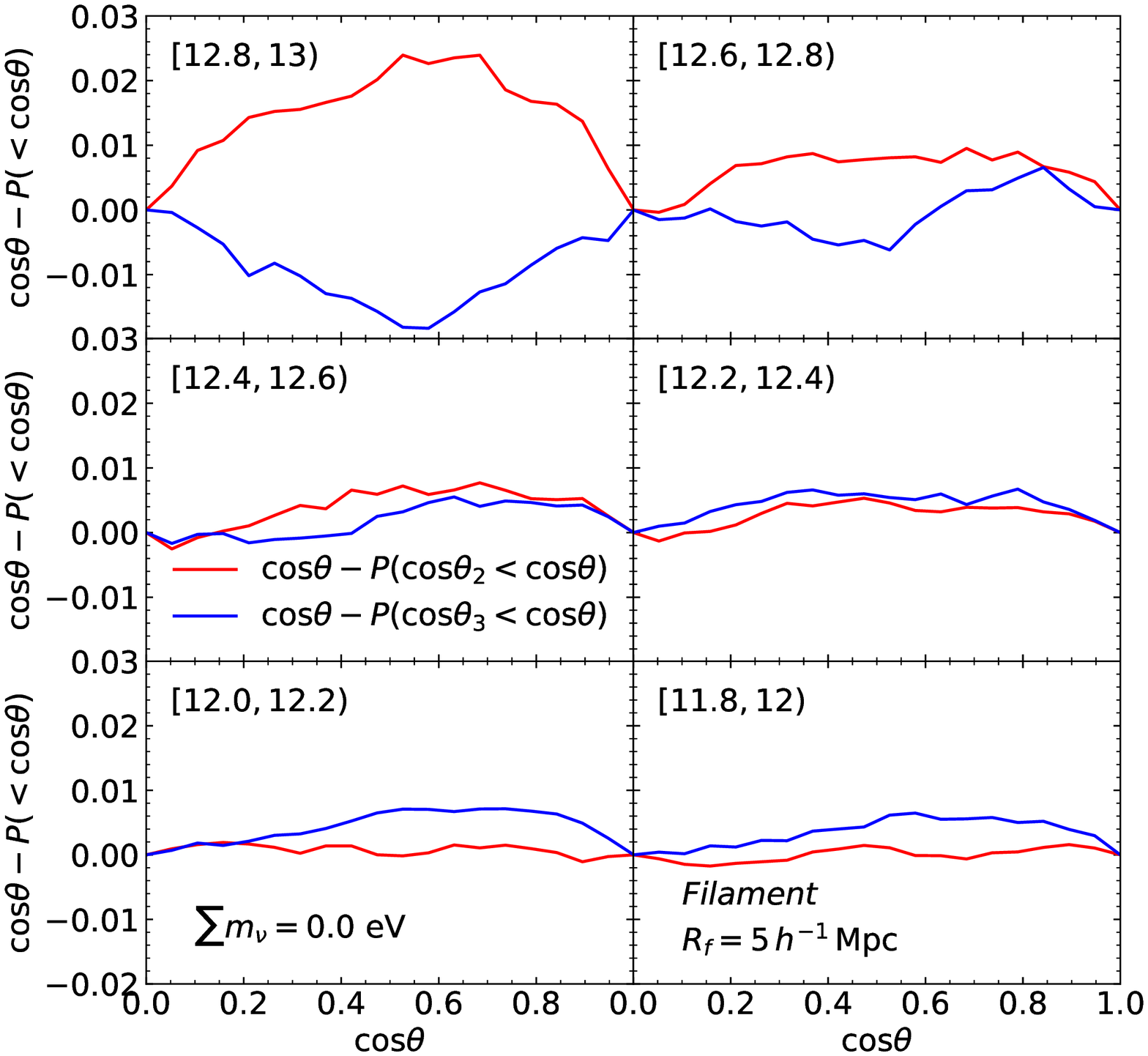}
\caption{Same as Figure \ref{fig:cbin_all_5_0.0} but in the filament environment.}
\label{fig:cbin_fil_5_0.0}
\end{center}
\end{figure}

\begin{figure}
\begin{center}
\includegraphics[scale=0.7]{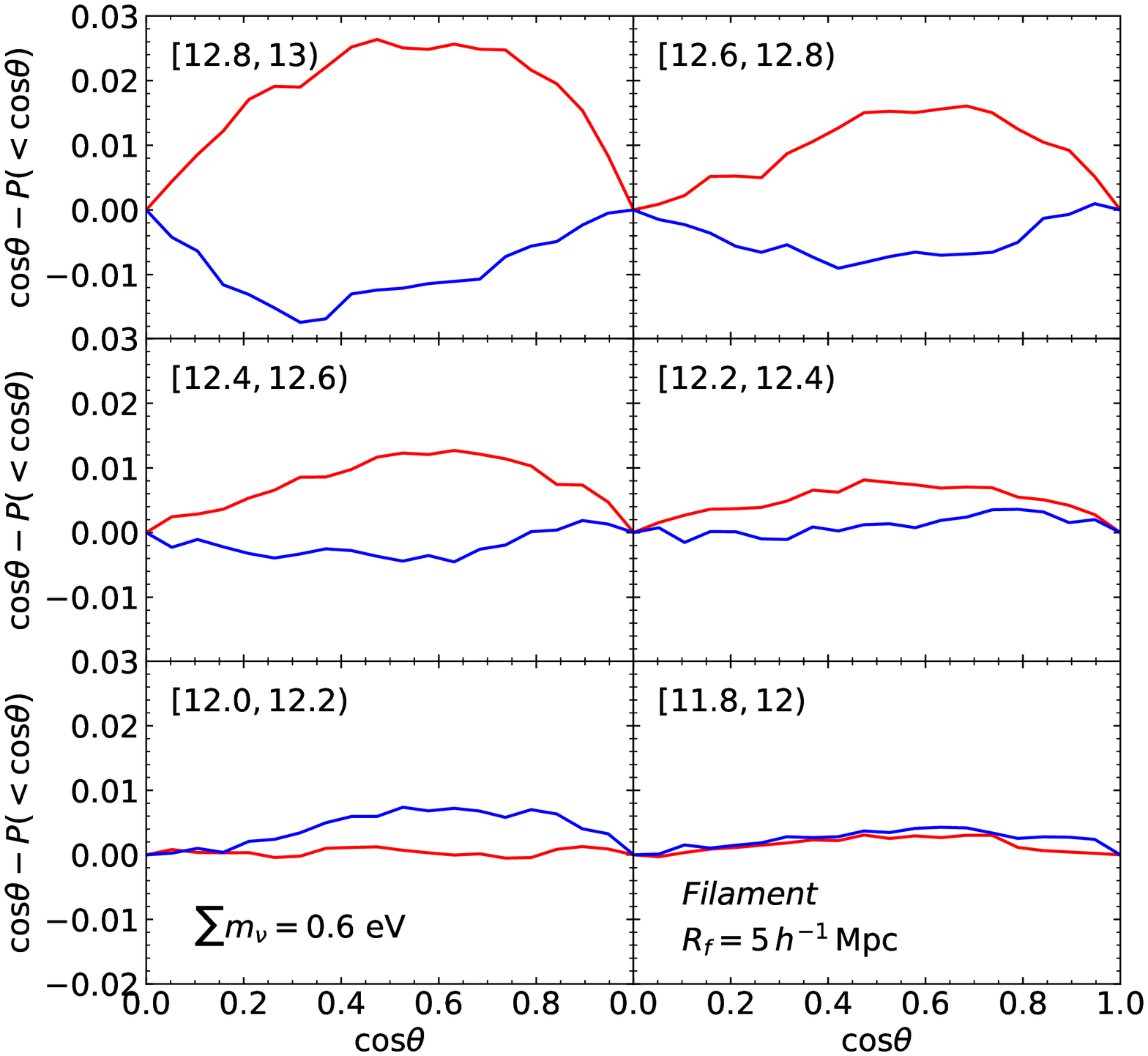}
\caption{Same as Figure \ref{fig:cbin_all_5_0.6} but in the filament environment.}
\label{fig:cbin_fil_5_0.6}
\end{center}
\end{figure}

Since the spin-flip phenomenon was known to be the most prominent in the filamentary environment \citep[][]{ara-etal07,cod-etal12,gan-etal18,lee19,kra-etal20}, 
we refollow the whole procedure but with only those halos located in the grid points at which the filament condition of 
$\lambda_{2}\ge 0, \lambda_{3}<0$ is satisfied \citep{hah-etal07a}. 
Figures \ref{fig:eali_fil_5}-\ref{fig:cbin_fil_5_0.6} plot the same as Figures \ref{fig:eali_5}-\ref{fig:cbin_all_5_0.6}, respectively, but using only the 
filament halos.   The condition for the occurrence of the spin flip, the drop of the confidence level for the rejection of the null hypothesis below $99.9\%$ 
is found to be satisfied at $\mlflip=(12.3\pm 0.1)$ and $\mlflip=(11.9\pm 0.1)$ for the cases of $\numass=0.0\ev$ and $\numass=0.6\ev$, respectively.  
The filament halos exhibit a larger difference in $\mflip$ between the two $\nulcdm$ models, which implies that the filaments are indeed optimal environment 
for the investigation of the $\numass$-dependence of $\mflip$.

\begin{figure}
\begin{center}
\includegraphics[scale=0.7]{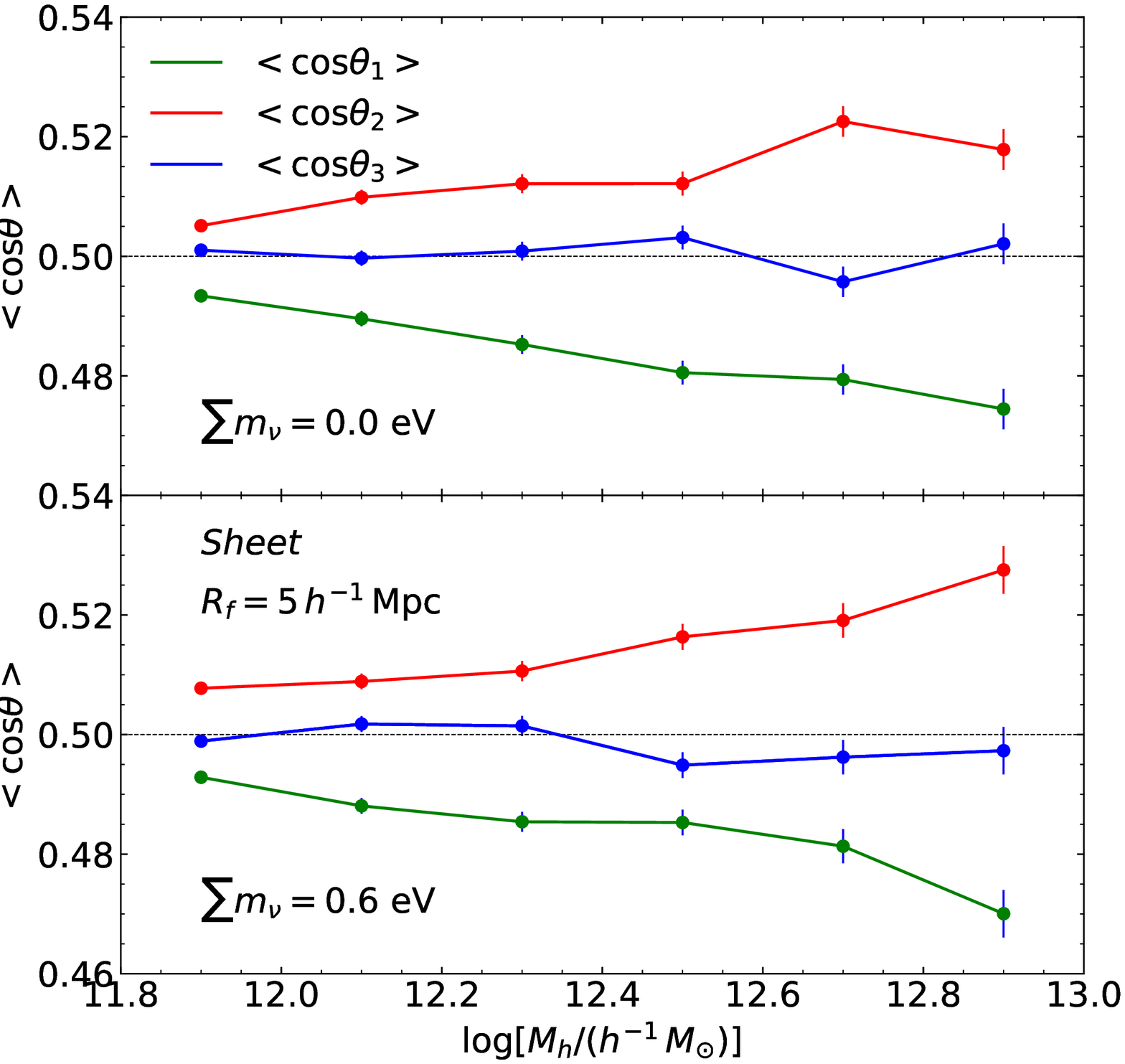}
\caption{Same as Figure \ref{fig:eali_5} but in the sheet environment.}
\label{fig:eali_sheet_5}
\end{center}
\end{figure}

In a similar manner, we also examine if the $\numass$-dependence of $\mflip$ can be found in the sheets ($\lambda_{1}>0, \lambda_{2}<0$) 
\citep{hah-etal07a}, the results of which are shown in Figure \ref{fig:eali_sheet_5}. As can be seen,  
in the sheets on the scale of $R_{f}=5\dunit$, we find no occurrence of the halo spin flips since 
$\mcthb>\mcthc\sim 0.5$ in the whole mass range for both of the $\nulcdm$ cosmologies. Note also that in the 
sheet environments on the scale $R_{f}= 5\,h^{-1}$Mpc, the intrinsic spin alignments of galactic halos with the tidal eigenvectors 
-for both of the models follow very well the predictions of the linear TTT, which is consistent with the observational result of 
\citet{lee-etal18}. No signal of the $\numass$-dependence of $\mflip$ is found even when $R_{f}$ varies from  $5\dunit$ to $10\dunit$.

\begin{figure}
\begin{center}
\includegraphics[scale=0.7]{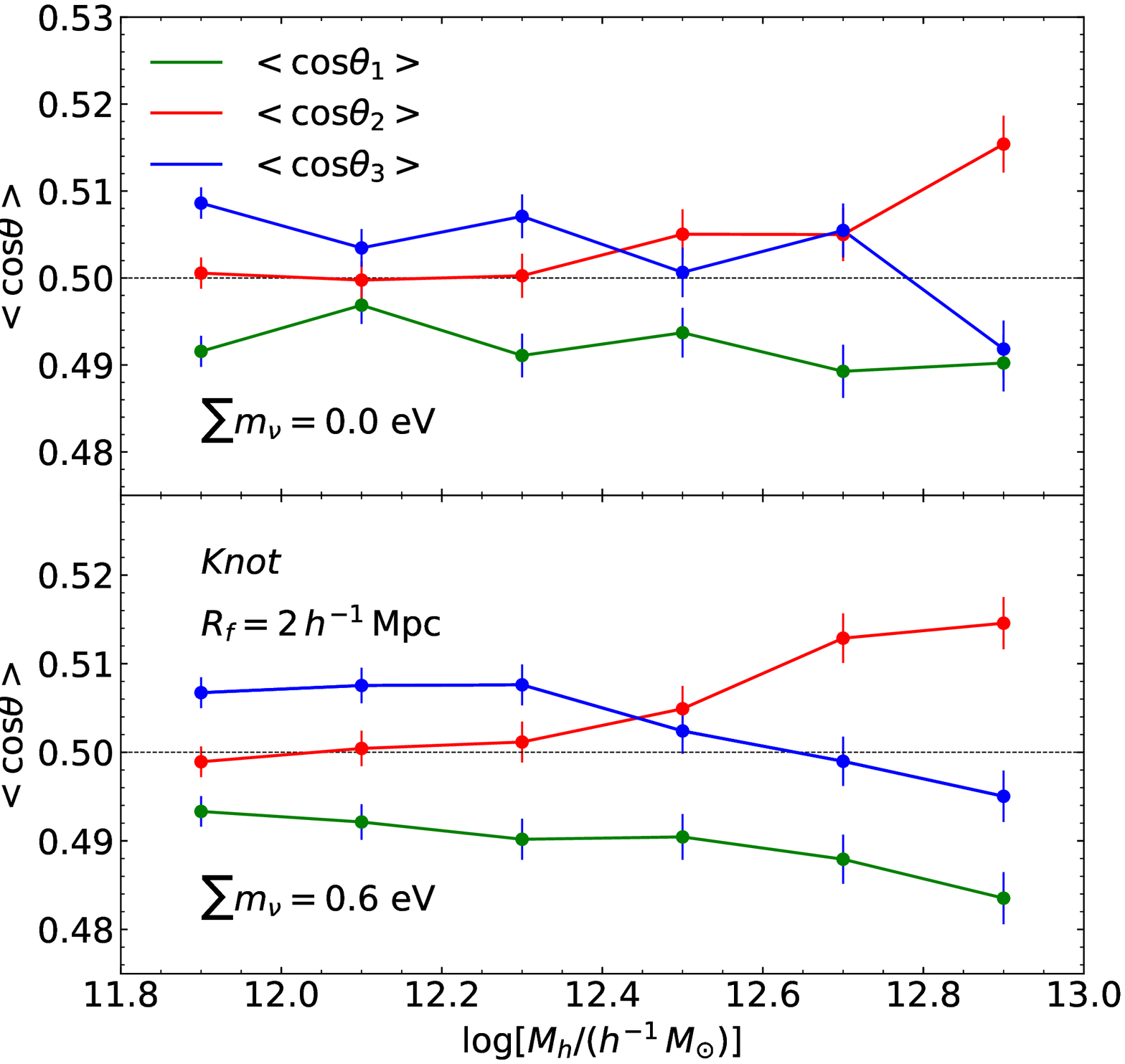}
\caption{Same as Figure \ref{fig:eali_fil_5} but in the knot environment on the scale of $R_{f}=2\dunit$.}
\label{fig:eali_knot_2}
\end{center}
\end{figure}

\begin{figure}
\begin{center}
\includegraphics[scale=0.7]{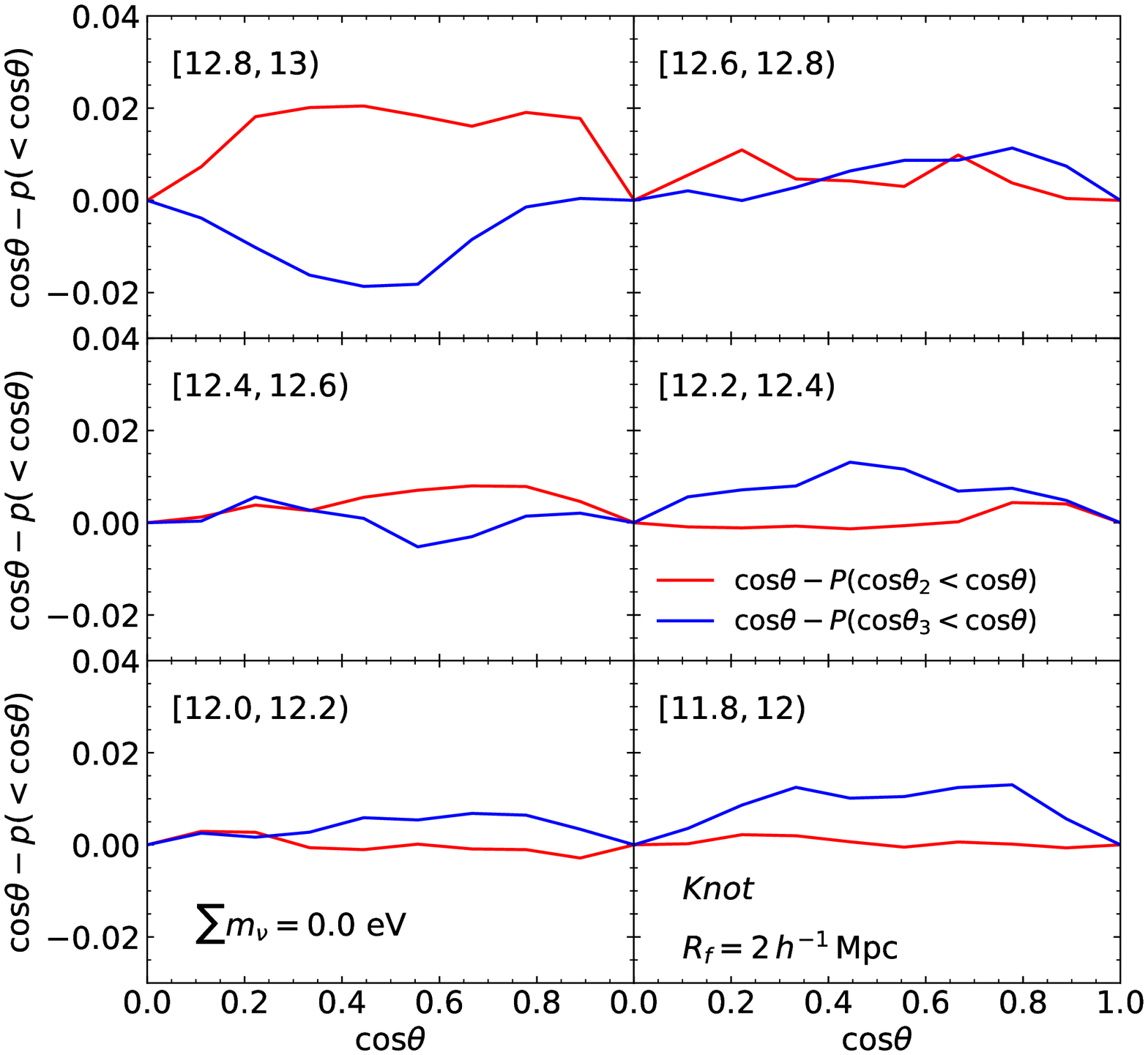}
\caption{Same as Figure \ref{fig:cbin_fil_5_0.0} but in the knot environment on the scale of $R_{f}=2\dunit$.}
\label{fig:cbin_knot_2_0.0}
\end{center}
\end{figure}

\begin{figure}
\begin{center}
\includegraphics[scale=0.7]{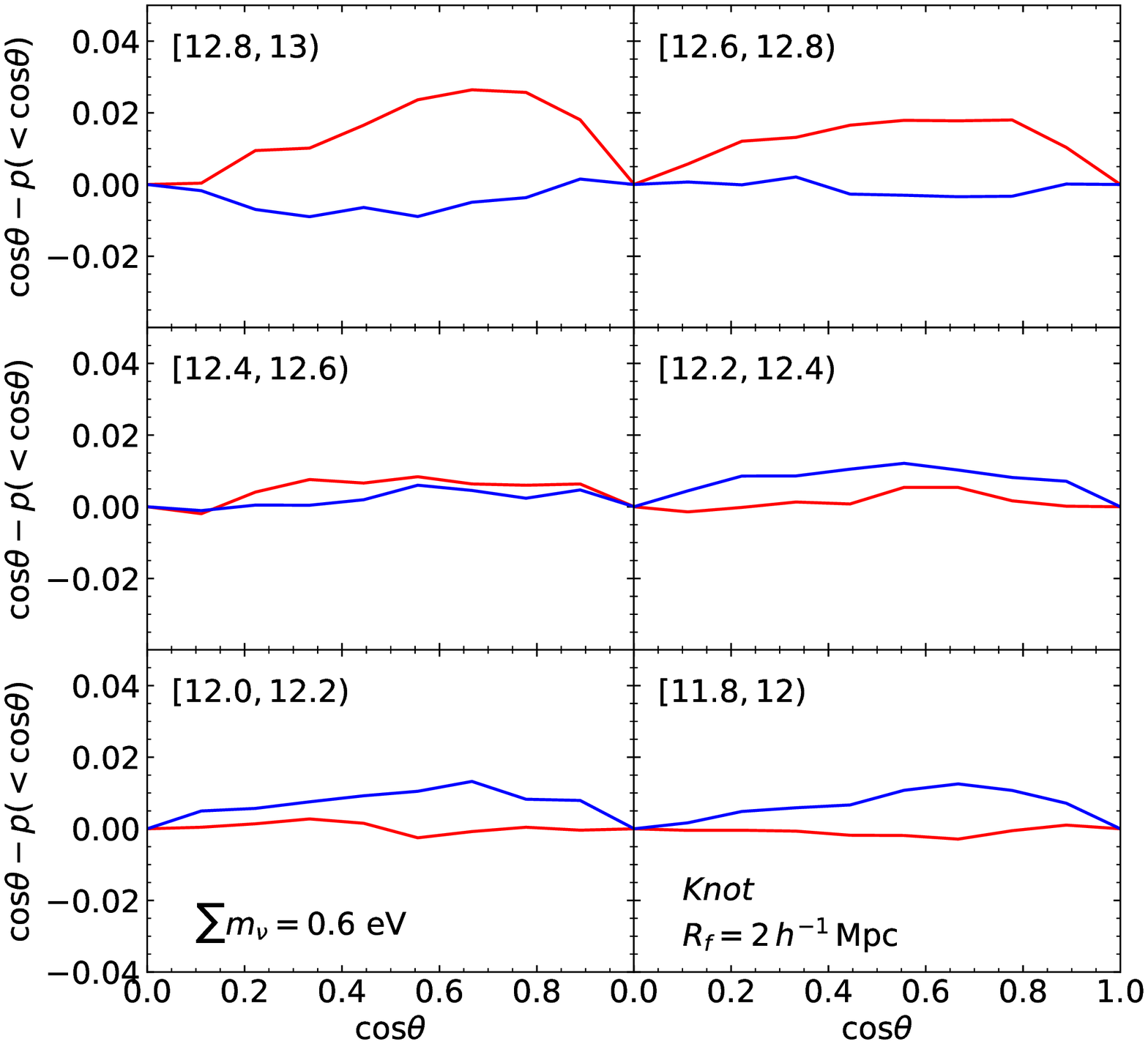}
\caption{Same as Figure \ref{fig:cbin_knot_2_0.0} but for the case of $\numass=0.6\ev$.}
\label{fig:cbin_knot_2_0.6}
\end{center}
\end{figure}

For the knot halos ($\lambda_{3}>0$), we detect a clear signal of the $\numass$-dependence of $\mflip$ on the scale of $R_{f}=2\,h^{-1}$Mpc, 
as shown in Figure \ref{fig:eali_knot_2}-\ref{fig:cbin_knot_2_0.6}.  The spin flips are found to occur at $\mlflip=12.7\pm 0.1$ and $12.5\pm 0.1$ 
for the cases of $\numass=0.0\ev$ and $0.6\ev$, respectively. 
The results from the knots on the larger scales of $R_{f}=5\,h^{-1}$Mpc and $10\,h^{-1}$Mpc as well as from the voids are found to carry large uncertainties 
due to poor number statistics and thus omitted here.  

Table \ref{tab:mflip} summarizes the logarithmic mass bins of $\log\!\mflip$ determined through the KS test according to our new definition for all of the different cases 
of the web type and $R_{f}$ considered for both of the $\nulcdm$ models.  It also lists the logarithmic mass bins of $\log\!\mflip^{\rm old}$ for comparison, where 
$\mflip^{\rm old}$ denotes the threshold mass for the occurrence of the halo spin flip defined in the conventional way using the criterion of $\mcthc=0.5$.

\section{Discussion and Conclusion}\label{sec:con}

Analyzing the numerical data from the $\nus$ \citep{liu-etal18} and exploring the intrinsic spin alignments of the galactic halos with the eigenvectors of the local tidal 
fields for $\nulcdm$ models that share the same initial conditions other than $\numass$, we have detected a clear trend that in the presence of 
more massive neutrinos, the spin flip occurs at a lower mass scale. The trend has been found to be the most prominent in the filaments, being robust against 
the scale variation. We interpret the detected $\numass$-dependence of $\mflip$ as an evidence for a retarding effect of massive neutrinos on the nonlinear 
evolution of the tidal eigen directions. 

While in the proto-galactic regime $\hbj$ is aligned with $\hbe_{2}$ of the linear tidal field \citep{LP00,mot-etal20}, 
the nonlinear evolution of the tidal field drives $\hbj$ to develop its alignment with $\hbe_{3}$ \citep{hah-etal07b,lib-etal13,lee19}. 
The threshold mass for the occurrence of the halo spin flip, $\mflip$, marks the mass scale at which the strength of the nonlinearly developed $\hbj$-$\hbe_{3}$ 
alignments overtakes that of the initially induced $\hbj$-$\hbe_{2}$ alignments. The faster the tidal fields evolve,  the larger the value of $\mflip$ is. 
The presence of more massive neutrinos which exerts stronger suppression of the density growths also retards more severely the nonlinear modification of the tidal 
eigenvectors from the initial principal directions, leading the galactic halos to retain better the initial memory of the $\hbj$-$\hbe_{2}$ alignments even in the mildly 
nonlinear regime, and consequently rendering the spin flip to occur at a lower mass scale.
Our result reveals the potential of $\mflip$ as a new probe of $\numass$ on the galactic halo scales, which can complement the conventional probes 
on the cluster halo scales.   

It is, however, worth discussing two limitations of the current analysis and how to improve them in the future prior to using this new probe in practice. 
First,  the current analysis has not taken into account the fact that the spin directions of the luminous galaxies are misaligned with 
those of the host DM halos \citep[e.g.,][]{hah-etal10}. Given that the galaxies exhibit a different tendency and strength of the spin alignments with the 
large-scale structures from those of their hosting DM halos \citep[e.g.,][]{dub-etal14,cod-etal18,gan-etal19,kra-etal20}, 
it will be of critical importance to investigate whether or not the luminous galaxies also show the same degree of the $\numass$-dependence of 
$\mflip$ by using hydrodynamics simulations performed for $\nulcdm$ models.

Second, our finding of the $\numass$-dependence of $\mflip$ has been obtained from the simulations that includes the relic neutrinos only at the level 
of the background, which implies that this new probe may not be free from the long-standing $\sigma_{8}$-$\numass$ degeneracy.  
Although the $\nulcdm$ model with $\numass=0.6\ev$ has the same amplitude of the primordial power spectra, $A_{s}$, as the $\Lambda$CDM cosmology 
with massless neutrinos, they differ in the derived values of $\sigma_{8}$ (Table \ref{tab:initial}), which should be at least partially contributed to their 
differences in $\mflip$. 
A more comprehensive study based on a simulation that incorporates nonlinearly the relic neutrinos \citep{zhu-etal14} will be required to investigate 
if and how $\mflip$ truly varies with $\numass$ on the nonlinear scales and to determine whether or not the $\numass$-dependence of $\mflip$ can 
break the $\sigma_{8}$-$\numass$ degeneracy. Our future work will be in this direction. 

\acknowledgments

We thank an anonymous referee for providing us very helpful suggestions. 
We thank the Columbia Lensing group for making their suite of simulated maps available at the website (http://columbialensing.org), and NSF for supporting 
the creation of those maps through grant AST-1210877 and XSEDE allocation AST-140041. We thank the New Mexico State University (USA) and Instituto de 
Astrofisica de Andalucia CSIC (Spain) for hosting the Skies \& Universes site for cosmological simulation products. We thank J.Liu for providing us with the snapshot 
data. We also thank the Lorentz Center of the Leiden University for the hospitality during the "Cosmic Web in the Local Universe" workshop where this work was initiated. 
J.L. and S.R. acknowledge the support by Basic Science Research Program through the National Research Foundation (NRF) of Korea 
funded by the Ministry of Education (No.2019R1A2C1083855) and also by a research grant from the NRF to the Center for Galaxy 
Evolution Research (No.2017R1A5A1070354). NIL acknowledges financial support of the Project IDEXLYON at the University of Lyon under the Investments for the Future 
Program (ANR-16-IDEX-0005). 
NIL also acknowledges support from the joint Sino-German DFG research Project ``The Cosmic Web and its impact on galaxy formation and alignment'' (DFG-LI 2015/5-1).

\clearpage
\clearpage
\clearpage
\clearpage
\clearpage
\clearpage
\clearpage
\clearpage
\clearpage
\clearpage
\clearpage
\clearpage
\begin{deluxetable}{cccccccc}
\tablewidth{0pt}
\setlength{\tabcolsep}{3mm}
\tablecaption{Initial conditions and the abundance of the galactic halos}
\tablehead{$\numass$ & $\Omega_{m}$ & $\Omega_{b}$ & $h$ & $n_{s}$ & $A_{s}$ & $\sigma_{8}$ & $N_{g}$ 
\\ $[\ev]$ & & & & &$[10^{-9}]$& &}
\startdata
$0.0$ & 0.3 & 0.046 & 0.7 & 0.97& 2.1 & 0.85 & 689654\\
$0.1$ & 0.3 & 0.046 & 0.7 & 0.97 & 2.1 & 0.83 & 680649\\
$0.6$ & 0.3 & 0.046 & 0.7 & 0.97 & 2.1 & 0.74 & 661030
\enddata
\label{tab:initial}
\end{deluxetable}
\clearpage
\begin{deluxetable}{ccccc}
\tablewidth{0pt}
\setlength{\tabcolsep}{3mm}
\tablecaption{}
\tablehead{$\numass$ & web type & $R_{f}$ & $\log(\mflip/\munit)$ & $\log(\mflip^{\rm old}/\munit)$ 
\\ $[\ev]$ & & [$\dunit]$ & & }
\startdata
$0.0$ & all & $5$ & $[12.2,\ 12.4]$  & $[12.6,\ 12.8]$\\
$0.6$ & all & $5$ &  $[12.0,\ 12.2]$& $[12.4,\ 12.6]$\\
\hline
$0.0$ & all & $10$ &  $[12.4,\ 12.6]$  & $[12.6,\ 12.8]$ \\
$0.6$ & all & $10$ &  $[12.2,\ 12.4]$  & $[12.6,\ 12.8]$ \\
\hline
$0.0$ & filament & $5$ &  $[12.4,\ 12.6]$  & $[12.6,\ 12.8]$ \\
$0.6$ & filament & $5$ &  $[12.0,\ 12.2]$  & $[12.2,\ 12.4]$ \\
\hline
$0.0$ & knot & $2$ &  $[12.6,\ 12.8]$  & $[12.6,\ 12.8]$ \\
$0.6$ & knot & $2$ &  $[12.4,\ 12.6]$  & $[12.6,\ 12.8]$ \\
\enddata
\label{tab:mflip}
\end{deluxetable}

\end{document}